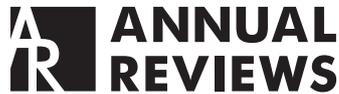

*Annual Review of Control, Robotics, and Autonomous Systems*

# Analysis and Control of Autonomous Mobility-on-Demand Systems

Gioele Zardini,[1,*] Nicolas Lanzetti,[2,*] Marco Pavone,[3] and Emilio Frazzoli[1]

[1]Institute for Dynamic Systems and Control, Department of Mechanical and Process Engineering, ETH Zürich, Zürich, Switzerland; email: gzardini@ethz.ch, efrazzoli@ethz.ch

[2]Automatic Control Laboratory, Department of Information Technology and Electrical Engineering, ETH Zürich, Zürich, Switzerland; email: lnicolas@ethz.ch

[3]Autonomous Systems Lab, Department of Aeronautics and Astronautics, Stanford University, Stanford, California, USA; email: pavone@stanford.edu

**Abstract**

Challenged by urbanization and increasing travel needs, existing transportation systems need new mobility paradigms. In this article, we present the emerging concept of autonomous mobility-on-demand, whereby centrally orchestrated fleets of autonomous vehicles provide mobility service to customers. We provide a comprehensive review of methods and tools to model and solve problems related to autonomous mobility-on-demand systems. Specifically, we first identify problem settings for their analysis and control, from both operational and planning perspectives. We then review modeling aspects, including transportation networks, transportation demand, congestion, operational constraints, and interactions with existing infrastructure. Thereafter, we provide a systematic analysis of existing solution methods and performance metrics, highlighting trends and trade-offs. Finally, we present various directions for further research.

# 1. INTRODUCTION

Countries worldwide are undergoing dramatic urbanization: Today, 55% of the world's population lives in urban areas, and the proportion is expected to reach 68% by 2050 (1). Urbanization leads to an increase in urban travel, and hence also in the externalities that this travel produces. Indeed, cities are considered responsible for 78% of the world's energy consumption and more than 60% of global emissions (30% of which is produced by transportation, in the United States) (2). Furthermore, congestion levels are increasing worldwide: In the United States, congestion causes 8.8 billion hours of extra travel time, equivalent to more than a week of vacation for the average US commuter, causing $180 billion of delay and fuel costs, which is close to 1% of the US gross domestic product (3).

In this context, private cars are considered an unsustainable solution for future personal mobility (4). Indeed, vehicles are typically overengineered (designed for long-distance highway travel but used mostly for low-speed urban trips) and underutilized (private cars are usually parked more than 90% of the time). The mobility-on-demand (MoD) concept has potential to alleviate these issues. In a MoD system, (light, electric) vehicles are placed throughout a city, and when customers want to reach a certain destination, they can walk to the nearest vehicle, pick it up, drive to the destination, and drop it off. MoD services directly target issues such as pollution and parking spaces (due to high utilization rates) and satisfy needs of personal mobility (4, 5).

However, MoD systems suffer from limitations, such as their tendency to become unbalanced due to unevenly distributed transportation demand. To overcome such issues, ride-hailing companies started offering a variation of MoD services, whereby passengers are assigned by a central manager to an on-demand driver, who then drives them to their destination. Although such services have become very popular [in New York City, ride-hailing companies increased their number of rides by 1,000% between 2012 and 2019 (6)], they exploit public resources (e.g., roads), require a large number of drivers to scale, and often lead to potentially disruptive consequences for the transportation system and for society at large (7, 8). Furthermore, their profitability is questionable.

Technological advances in the field of autonomous driving over the past few decades offer a new mobility paradigm: autonomous mobility-on-demand (AMoD). An AMoD system consists of a fleet of autonomous vehicles (AVs) that pick up passengers and transport them to their destination. A manager controls the fleet by simultaneously assigning passengers to AVs, routing them, and rebalancing the fleet by relocating customer-free AVs to realign their geographical distribution with transportation demand. One could argue that MoD systems in which drivers precisely follow instructions from a fleet manager could provide the same service as AMoD systems. So what justifies AMoD over MoD?

AMoD promises two main benefits. First, it increases the supply of drivers to match increasing demand with good quality of service. Indeed, American drivers spend approximately 1 h per day driving (9), equivalent to one-ninth of a work day, considering commute time to be productive time. Assuming driving is allocated to the same group of people (i.e., ignoring demand peaks), one out of nine workers should be a MoD driver. Additionally, customer-free trips, for which a common lower bound is 30% (e.g., based on Singapore data), increase the proportion to one out of seven, hindering the scalability of MoD systems. Second, AMoD drastically reduces transportation costs (10, 11). For instance, Becker et al. (11) estimated that electrification and automation would reduce the costs of taxis by more than 70% in Austin and 85% in Zürich.

Analytically and computationally, the operation of AMoD systems is similar to that of taxi systems, with one key difference: A taxi system operator cannot enforce assignments of passengers to drivers and balanced vehicle distributions because drivers have the final say. Typically, taxi drivers act selfishly to maximize their own profit, while AVs can act cooperatively. In AMoD systems, central coordination schemes can minimize externalities by optimizing routing, rebalancing, charging schedules, and so on. Furthermore, customers do not need to drive and can thus better use commute time. Driven by such promising benefits, various companies started to develop the technology (approximately 30 AV companies just in California), and some are already providing mobility services. Examples include Motional, whose AVs have provided more than 100,000 rides in Las Vegas (12), and Waymo, whose AVs have driven more than 20 million miles (13).

Overall, the novelty of AMoD systems calls for methods for analysis and control. The challenges are sociotechnical, ranging from technical ones (e.g., reaching sufficient levels of autonomy and defining operational regulations and liabilities) to political and economic ones (e.g., taxation policies, privacy, and fairness) (14). Questions of interest include which kind of autonomy and what levels of safety and reliability are needed for AVs to operate in cities, how many AVs are needed to guarantee a desired quality of service, how AMoD systems will interact with existing mobility systems, whether AMoD systems will reduce congestion, and whether we can use AMoD systems to steer urban mobility toward a sustainable future.

Numerous reviews have been published on aspects relevant to the analysis and control of AMoD systems, ranging from the vehicle routing problem (15–20) to shared (autonomous) mobility (21–27). However, to the best of our knowledge, no review has focused specifically on the analysis and control of AMoD systems, identifying problem settings, models, solution approaches, and future challenges.

In this article, we provide a synoptic review of methods for the analysis and control of AMoD systems. Specifically, our contribution comprises four parts. First, we identify problem settings for analysis and control of AMoD systems. Second, we present mathematical models, including models for the transportation network, transportation demand, congestion, operational constraints, and interactions with existing infrastructure. Third, we review solution methods, together with performance metrics and case studies. Our analysis highlights patterns and instances yet to be solved. Finally, we provide an outlook on future research.

## 2. ANALYSIS AND CONTROL OF AUTONOMOUS MOBILITY-ON-DEMAND SYSTEMS

Before introducing mathematical tools to reason about AMoD systems, we clarify why we need to analyze and control them in the first place. Essentially, there are two aspects:

- Operational policies: The deployment of AMoD systems calls for real-time operational policies. Questions of interest include which AV should pick up a customer requesting a ride, how AVs should be routed throughout the city, what customer-free AVs should do, and whether multiple requests can be simultaneously accommodated by sequentially picking up customers and driving them to their destinations.
- Planning studies: Operations of AMoD systems rely on planning. Questions to answer include how many AVs are needed to provide a given quality of service, where charging stations should be located, and how AVs will affect existing mobility systems.

Addressing these questions requires two ingredients: a mathematical model of the AMoD system and a solution approach (**Figure 1**). We first review problems related to the analysis and control of AMoD systems, distinguishing operational policies (Section 2.1) and planning (Section 2.2). In Section 3, we investigate models for AMoD systems, describing models for transportation networks, transportation demand, congestion, operational constraints, and interactions. Finally, in

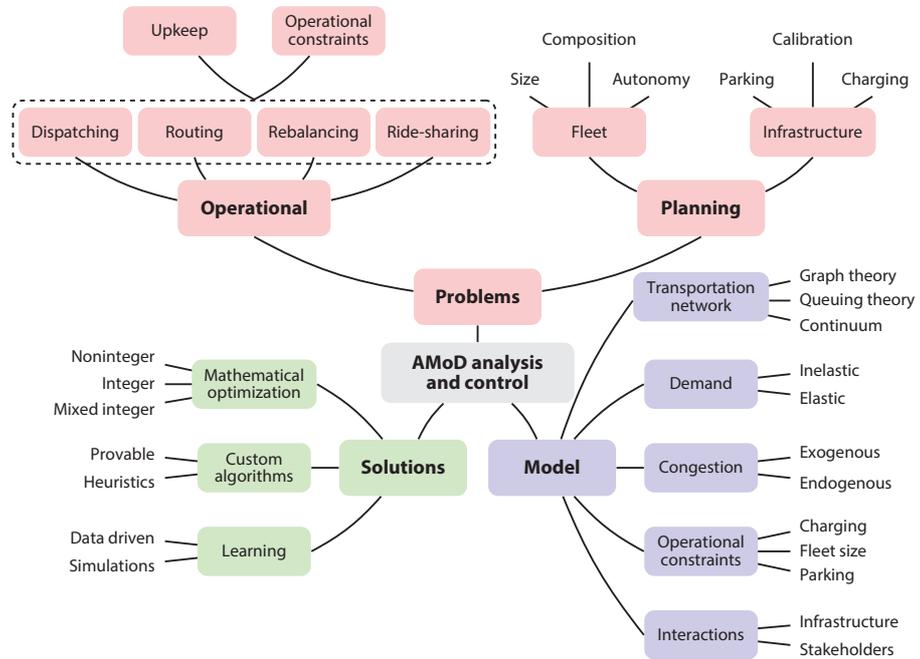

**Figure 1**

Dependency diagram for the analysis and control of autonomous mobility-on-demand (AMoD) systems, featuring mathematical models, problems, and solution approaches.

Section 4, we present solution approaches for operational matters and planning problems as well as related performance metrics.

### 2.1. Operational Policies

We distinguish four operational tasks: dispatching, routing, rebalancing, and ride-sharing (**Figure 2**). Additionally, each task could account for the upkeep of AVs, which includes scheduling of charging, calibration, parking, and general maintenance routines.

**2.1.1. Dispatching.** Given a set of customer requests, the fleet manager needs to assign them to available AVs (**Figure 2a**). At their core, dispatching problems are assignment problems, which have been well studied in operations research (28–31). Typically, when solving such problems, one seeks a system optimum that maximizes fleet performance.

**2.1.2. Routing.** After being dispatched, AVs need to be (dynamically) routed to the assigned customers and then to the destinations of their trips (**Figure 2b**). Routes are typically chosen to optimize specific metrics (e.g., travel time and travel distance), possibly considering road congestion. Such problems belong to the family of vehicle routing problem, in particular dynamic pickup and delivery problems (reviewed in 18, 32–34).

**2.1.3. Rebalancing.** Rebalancing problems deal with repositioning of customer-free AVs (35) (**Figure 2c**). Policies aim to maximize the responsiveness of AMoD systems to new requests while minimizing the imbalance of the AV fleet caused by asymmetries in transportation demand.

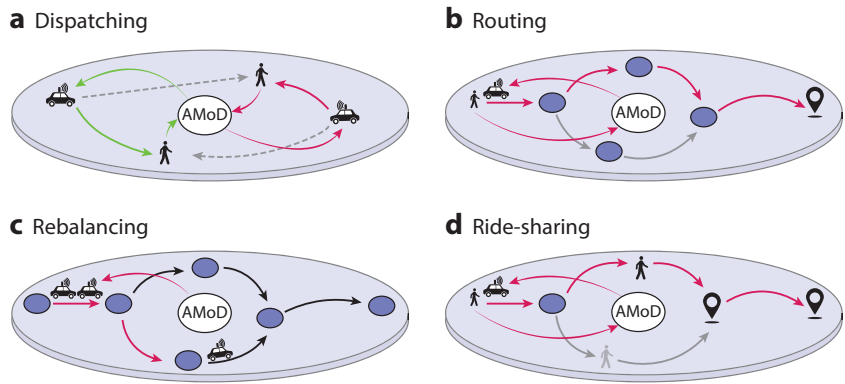

**Figure 2**

Operational policies for autonomous mobility-on-demand (AMoD) systems. An AMoD fleet manager gathers information from customers and autonomous vehicles, processes it, and produces actionable policies for the vehicles.

Rebalancing techniques usually consist of two steps: an estimate of future transportation demand and an efficient redistribution of customer-free AVs.

**2.1.4. Ride-sharing.** Ride-sharing refers to the process of simultaneously serving multiple, unrelated customer requests with a single AV (**Figure 2d**). This additional feature complicates the solution of the aforementioned problems: Ride-sharing policies need to provide AV assignments and routes dealing with multiple pickup and drop-off locations while maximizing fleet performance and not causing a deterioration in quality of service (36–38).

## 2.2. Planning Autonomous Mobility-on-Demand Operations

Numerous planning problems have been investigated, which we cluster in three categories.

**2.2.1. Design of autonomous vehicle fleets.** The design of AV fleets impacts the performance of AMoD systems. Design problems include fleet sizing (39–43), fleet composition (e.g., vehicle capacity and propulsion) (44), and autonomy (45, 46). For instance, Spieser et al. (39) solved two instances of fleet sizing: minimum fleet sizing and performance-driven fleet sizing. The former seeks the minimum number of AVs in a fleet to keep the number of unmet demands bounded [92,693 AVs in Singapore (39)], while the latter quantifies the number of AVs for a certain quality of service. Fleet sizing also informs auxiliary operations related to AMoD systems. For instance, in low-demand regimes, unnecessary AVs could be sent to charging and calibration facilities or provide auxiliary services, such as package delivery (47).

**2.2.2. Interactions with existing infrastructure.** AMoD systems operate while using public and private resources, such as space for parking and idling, roads, and electricity. Therefore, planning studies also focus on interactions between AMoD systems and existing infrastructure. Aspects of interest include the allocation of parking space (48, 49), the optimal placement of charging stations, charge scheduling for electric AVs (50–54), contributions to congestion (55), joint operation with other modes of transportation (56, 57), autonomous calibration (58), smart infrastructure (59), and the achievable performance of the overall mobility system (45).

### 2.2.3. Interactions among stakeholders.
AMoD systems are part of a large ecosystem of mobility providers that includes public transportation systems and private companies. Importantly, studies investigate the interplay among AMoD systems, other mobility solutions, and regulatory agencies (e.g., municipalities) (60, 61), focusing on pricing strategies, incentives, and tolls (51, 62–65). While constituting an exciting research direction, the study of interactions among stakeholders is not the focus of this article. Nevertheless, we discuss the topic when posing future challenges in Section 5.

## 3. MATHEMATICAL MODELING OF AUTONOMOUS MOBILITY-ON-DEMAND SYSTEMS

In this section, we review models for AMoD systems, including modeling approaches for the transportation network, AVs, transportation demand, congestion models, operational constraints, and interactions with existing infrastructure. We focus the exposition on mathematical models. We discuss simulation-based and empirical approaches, often combined with machine learning, in Section 4.

### 3.1. Transportation Network and Autonomous Vehicles

Models of transportation networks and AVs can be clustered in two categories:

- Time-varying models consider both the spatial and the temporal dimensions of the transportation system. Typically, authors focus on a specific geographical area (e.g., Manhattan) and a specific time window (e.g., one day). This approach is common when modeling microscopic phenomena or making operational decisions but usually leads to more complicated models (e.g., larger graphs).
- Time-invariant models assume a transportation system in steady state, where one neglects the temporal dimension, by assuming a specific time window and treating it as a single time unit (e.g., considering the number of AVs per unit time). This assumption is reasonable if requests change slowly compared with the average travel time of individual trips, as is often observed in densely populated urban environments (56, 66). This approach is popular in mesoscopic and macroscopic studies, as it simplifies the setting while capturing the main aspects of the problem.

In the following, we present mathematical models and relate them to the time-varying and time-invariant approaches (**Figure 3**).

#### 3.1.1. Graph-theoretic models.
The transportation network is modeled as a directed graph $\mathcal{G} = \langle \mathcal{V}, \mathcal{E} \rangle$, where $\mathcal{V}$ is the set of vertices and $\mathcal{E} \subseteq \mathcal{V} \times \mathcal{V}$ is the set of edges (**Figure 3a**). A node $v \in \mathcal{V}$ represents a location, such as a station or an urban area, and an edge $\langle v_1, v_2 \rangle \in \mathcal{E}$ represents a road (or a combination of roads) from location $v_1$ to location $v_2$. Each edge is usually associated with metrics such as travel time $t : \mathcal{E} \to \mathbb{R}_{\geq 0}$ and distance $d : \mathcal{E} \to \mathbb{R}_{\geq 0}$ [e.g., $t_{ij} := t(i,j)$ is the travel time of edge $\langle i,j \rangle$].

When using graph-theoretic models, one often adopts an (Eulerian) fluid-dynamic approach: AVs and customers are represented not individually but as (possibly noninteger) flows between nodes. In this setting, time-invariant models usually reduce to static network flow problems (for details, see 67). Dynamic models, by contrast, are studied via differential equations (35, 68–71) or via time-expanded graphs (72). Differential equations capture the dynamic behavior of AVs and customers on the graph. For instance, Pavone et al. (35) showed that, by flow conservation, the number of AVs $v_i(t)$ and the number of customers $c_i(t)$ at node $i$ obey the nonlinear time-delayed

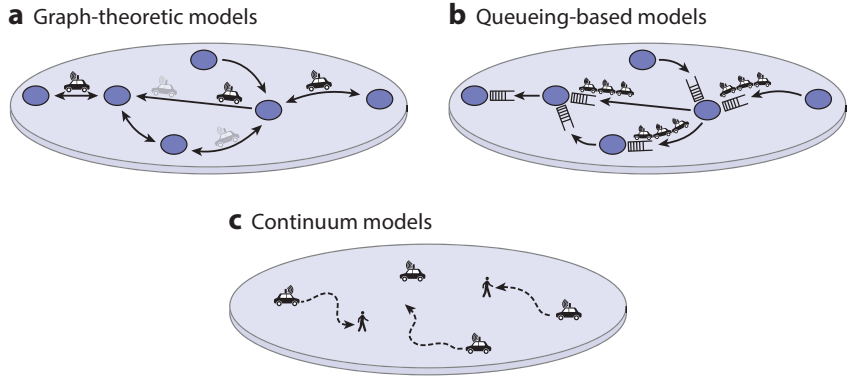

**Figure 3**

Models for a transportation network and autonomous vehicles.

differential equations

$$\dot{v}_i(t) = -\lambda_i H(v_i(t)) + (\lambda_i - \nu_i) H(v_i(t)) H(c_i(t))$$
$$+ \sum_{j \neq i} p_{ij} \left( \lambda_j H(v_j(t - t_{ji})) - (\lambda_j - \mu_j) H(c_j(t - t_{ji})) H(v_j(t - t_{ji})) \right)$$
$$- \sum_{j \neq i} \alpha_{ij} H(v_i(t)) + \sum_{j \neq i} \alpha_{ji} H(v(t - t_{ji})),$$
$$\dot{c}_i(t) = \lambda_i (1 - H(v_i(t))) + (\lambda_i - \nu_i) H(v_i(t)) H(c_i(t)),$$

1.

where $\lambda_i$ and $\mu_i$ are the rates of arrival and departure of customers at node $i$, $\alpha_{ij}$ is the rate of rebalancing AVs from node $i$ to node $j$, $p_{ij}$ is the probability that a customer at node $i$ wants to travel to node $j$, and $H(x)$ is the Heaviside function [$H(x) = 1$ if $x > 0$, $H(x) = 0$ otherwise]. Time-expanded graphs, by contrast, are a natural generalization of directed graphs for time-varying transportation networks. Herein, nodes are spatiotemporal and consist of a geographical location and a time instant: An AV is at $n = \langle v, t \rangle$ if it is at the physical node $v$ at time $t$ (73). Accordingly, an edge between $n_1 = \langle v_1, t_1 \rangle$ and $n_2 = \langle v_2, t_2 \rangle$ exists if and only if $v_2$ can be reached from $v_1$ in the time interval $t_2 - t_1$.

While most works adopt a fluidic approach, graph-theoretic models can be combined with vehicle-centric models, whereby AVs and customers are modeled individually (74–76). For instance, James & Lam (75) introduced a binary decision variable $x_{ij}^k \in \{0, 1\}$ to indicate whether AV $k$ is traversing edge $\langle i, j \rangle$ in order to optimize AV routes. Nonetheless, vehicle-centric models sometimes ignore the transportation network and model only AVs and customers (77–79). In the simplest setting, dispatching of AVs is optimized via binary decision variables $y_{ij} \in \{0, 1\}$ that take value 1 if and only if AV $i$ is assigned to customer $j$.

**3.1.2. Queuing-theoretic models.** Queuing theory deals with studies of waiting lines. In AMoD systems, trips are modeled as queues between locations (**Figure 3b**), allowing one to analyze characteristics such as the availability of AVs (i.e., the probability that at least one AV is available), which is closely related to waiting time. Formally, consider $N$ stations placed at specific geographical locations and $m$ AVs providing mobility service (80–82). Customers arrive at each station according to an exogenous stochastic process and select a destination with some probability (for more details on demand models, see Section 3.2). If AVs are available at a station, customers travel to their respective destinations; otherwise, they leave the system (so-called passenger loss). Travel times between stations are also stochastic and are usually modeled as exponentially distributed random variables. When designing policies for AMoD systems, one analyzes and regulates how AVs move from one queue to another.

These models are often deployed in time-invariant settings. The case of infinitely large fleets ($m \to \infty$) has been of particular interest to derive theoretic insights. For instance, George (80) showed that, for a fleet of $m$ AVs, the availability $A_i(m)$ of AVs at node $i$ satisfies

$$\lim_{m \to \infty} A_i(m) = \frac{\lambda_i/\pi_i}{\max_{i \in \{1,\ldots,N\}} \lambda_i/\pi_i},$$

where $\lambda_i$ is the average arrival rate at station $i$ and $\pi_i$ is the relative throughput at station $i$. These insights formally quantify the imbalance of the AMoD system and allow one to rigorously reason about rebalancing policies (82).

**3.1.3. Continuum models.** The transportation network is modeled as a compact domain $\Omega \subset \mathbb{R}^2$ of the Euclidean plane (**Figure 3c**). AVs are free to move on $\Omega$ and are typically assumed to have a constant speed. Abstracting away the topology of the road network substantially simplifies the analysis and synthesis of control strategies for AMoD systems (83), allowing one to derive analytical expressions for relevant operational parameters (18). For instance, Treleaven et al. (84) proved that for particular origin–destination distributions, the distance driven by customer-free AVs is lower bounded by the earth mover's distance (85) and provides sufficient and necessary conditions for load balancing of transportation systems. Recently, researchers also modeled AMoD systems using stochastic differential equations (86). Here, the stochasticity of the system dynamics is driven by nonstationary demand and vehicle arrival processes, allowing one to split operational problems into an average process (which can be estimated from historical data) and an unknown disturbance.

### 3.2. Transportation Demand

Transportation demand (henceforth referred to as demand) models the willingness of customers to travel (e.g., for business or leisure). We distinguish between inelastic and elastic demand. A demand model is inelastic if demand is independent of the operation of the AMoD system. It is elastic if demand is affected by the operation of the AMoD system.

**3.2.1. Inelastic demand.** Inelastic demand models can be classified as follows:

- Deterministic demand models assume a fixed (possibly time-varying) demand distribution, often derived from historical data. Such models are typically used along with graph-theoretic models, whereby demand is modeled via origin–destination pairs or via arrival and departure rates at each node. For instance, origin–destination pairs are specified by a pair of nodes $\langle o, d \rangle \in \mathcal{V} \times \mathcal{V}$ (for time-invariant settings) or by a pair of nodes and a departure time $\langle o, d, \tau \rangle \in \mathcal{V} \times \mathcal{V} \times \mathbb{R}_{\geq 0}$ (for time-varying settings) (45, 73).
- Stochastic demand models are ubiquitous in queuing-theoretic and continuum models. Often, they result in spatial (or spatiotemporal, in time-varying settings) stochastic processes. For instance, in queuing-theoretic models, customers are assumed to arrive at each station $i$ according to a time-invariant Poisson process and select a destination $j$ with some probability $p_{ij} \in [0, 1]$ and $\sum_{j=1}^{N} p_{ij} = 1$ for all $i$ (80–82).

**3.2.2. Elastic demand.** In real transportation systems, demand is elastic: A more efficient mobility service could induce demand, and conversely, a less efficient mobility service could decrease it. Herein, demand is assumed to be adversely affected by travel time. For dispatching problems, assignments under elastic demand can be computed via mathematical optimization, whereby the cost function combines congestion and disutility demand (see 87–90 and references therein). As shown by this review, researchers have focused on inelastic demand models; elastic demand, thus, represents an area for future research.

### 3.3. Congestion

Congestion accounts for delays in transportation systems. We distinguish between two types of congestion models: exogenous and endogenous. A congestion model is exogenous if congestion is a fixed, possibly time-varying, known parameter (affecting travel time), independent of the operation of the AMoD system. Conversely, it is endogenous if it is affected by the operation of the AMoD system.

**3.3.1. Exogenous congestion.** Exogenous congestion models correct (free-flow) travel time to account for congestion caused by all other users on the road. Estimates often stem from historical traffic information. These models are justified for relatively small fleet sizes: A small AMoD system will not significantly increase congestion on the roads (5). Exogenous congestion leads to more tractable models of transportation systems. For instance, when making operational decisions, one does not need to consider how policies affect travel time. Yet congestion-aware operations of AMoD systems promise to mitigate congestion (5).

**3.3.2. Endogenous congestion.** Endogenous congestion models are found in queuing-theoretic and graph-theoretic models. In queuing-theoretic models, congestion is typically considered through capacity constraints on the queues. Graph-theoretic models offer more flexibility: Congestion can be modeled via capacity constraints on the edges (so-called threshold models) or via increased travel times. In time-invariant network flow models, capacity constraints upper bound the total flow of vehicles, including exogenous traffic $u_{ij}$ (i.e., vehicles already on the road) and AVs $f_{ij}$, on edge $\langle i, j \rangle$ (e.g., $u_{ij} + f_{ij} \leq 10$ AVs per second). For increased travel times, instead, one assumes that travel time $t_{ij}$ on edge $\langle i, j \rangle$ depends on the road usage of all users of the transportation system, including the AMoD system itself. Formally, travel delays are expressed by a nonnegative strictly increasing function $\Phi : \mathbb{R}_{\geq 0} \to \mathbb{R}_{\geq 0}$ mapping the total number of vehicles on an edge to a delay: Given a flow of AVs $f_{ij}$ on edge $\langle i, j \rangle$, travel time results from

$$t_{ij} = t_{ij}^{\text{F}} \cdot \Phi\left(\frac{f_{ij} + u_{ij}}{c_{ij}}\right),$$

where $c_{ij}$ is the nominal road capacity and $t_{ij}^{\text{F}}$ is the free-flow travel time.[1]

### 3.4. Operational Constraints

To analyze and control AMoD systems, one should consider the operational limits of AVs. Here, we present the constraints studied in the literature.

---

[1] For instance, the Bureau of Public Roads function reads $\Phi(x) = 1 + 0.15x^4$ (91, 92).

**3.4.1. Fleet size.** Fleet size constraints account for the limited number of AVs composing a fleet. We identify three instances of fleet size constraints. First, fleet size limitations are sometimes naturally embedded in the model of the transportation system, either because it accounts for every single AV (74, 75, 79) or because it is based on queuing theory (93). For instance, a queuing-theoretic model consists of $N$ stations and $m$ vehicles; thus, fleet size is implicitly specified by the model of the transportation network and AVs. Second, fleet size sometimes results from the initial condition of a dynamic model of the transportation system (35, 94). For instance, Pavone et al. (35) showed that, in the setting of Equation 1, the number of AVs is constant [i.e., $\frac{d}{dt} \sum_i v_i(t) = 0$]. Hence, fleet size results from the initial condition $\sum_i v_i(0)$. Third, fleet size can be imposed via hard constraints, especially in time-invariant network flow models, where the number of AVs can be estimated by multiplying the sum of the customer and rebalancing flows with travel time (45). Finally, we remark on two aspects of these constraints: First, each type of constraint can be readily generalized to mixed fleets by imposing limitations on each vehicle class (95), and second, fleet sizing represents not only an operational constraint but also a design parameter for planning studies (43, 45).

**3.4.2. Charging.** Another frequent constraint relates to operational limits of AVs. Considering ongoing electrification efforts, research has focused mainly on electric AVs and, thus, on charging constraints (52). Specifically, we distinguish two modeling approaches. The first one constrains the operation of AVs, either through charging- or refueling-related delays in travel time or through upper bounds on trip duration (74). In the second one, the battery of electric AVs has a dynamic state of charge, and operational policies are designed to optimally allocate charging schedules (53, 54, 68, 76, 96–98). Finally, researchers also consider the bidirectional interplay between transportation systems and power grids (54, 62).

**3.4.3. Parking.** The impact of AVs on parking spaces (and vice versa) is a relevant aspect for the management of transportation systems (49, 99). Parking policies typically relocate idling AVs while not causing deterioration in metrics such as waiting times, distance traveled, and congestion. In graph-theoretic models, effects of parking are captured by additional costs when idling at a node (price of parking) (11, 100, 101) or via upper bounds on the number of AVs at a node (capacity of parking lots) (48, 69, 70, 74). In queuing-theoretic models, stations are usually assumed to have an infinite number of parking spaces (82). Extensions to capacitated queues were discussed by Zhang (81).

### 3.5. Interactions

Interactions with existing infrastructure, such as public transportation systems (56) and power grids (54), as well as interactions among stakeholders in mobility ecosystems (64, 65), such as municipalities and other mobility providers, are relevant modeling aspects. For instance, the analysis of intermodal mobility systems (whereby an AMoD system provides service jointly with public transport) requires modeling both the AMoD system and the public transportation network (56). Formally, the network of each mode of transportation can be modeled as in Section 3.1.1, and networks are connected via so-called switching arcs. For simplicity, consider a road network $\mathcal{G}_R = \langle \mathcal{V}_R, \mathcal{E}_R \rangle$ and a subway network $\mathcal{G}_S = \langle \mathcal{V}_S, \mathcal{E}_S \rangle$. Switching edges (representing connections between roads and subway stations) are described by a set of edges $\mathcal{E}_C \subseteq \mathcal{V}_R \times \mathcal{V}_S \cup \mathcal{V}_S \times \mathcal{V}_R$, characterized by additional properties (e.g., walking time from a road to a subway station). The resulting network is $\mathcal{G} = \langle \mathcal{V}_R \cup \mathcal{V}_S, \mathcal{E}_R \cup \mathcal{E}_S \cup \mathcal{E}_C \rangle$ (for an extensive treatment, see 45). This model can be used to assess benefits of intermodality (56) and to devise operational policies (72).

## 4. SOLUTION APPROACHES

Recall the two classes of problems introduced in Section 2:

- Operational policies are designed to control AMoD systems in real time. Given a stream of requests, the system must be able to assign them to available AVs, fulfill them, and rebalance the system, while guaranteeing fleet performance.
- Planning studies address operational problems to investigate the design of AMoD systems, interactions with existing infrastructure, and interplay among stakeholders in the transportation system.

Performance metrics and optimization objectives include revenue and costs, waiting and travel time, travel distance, demand satisfaction, fleet size, and social welfare. We review solution methods (Section 4.1) and tools for simulations (Section 4.2) from both operational and planning perspectives. We classify solution approaches in **Table 1**. We consider modeling aspects of the transportation network (continuum, fluidic, queuing-theoretic, or vehicle-centric), operational constraints (fleet size, limited AV autonomy, and parking), upkeep (solutions for scheduling problems related to charging of AVs and dynamic parking of AVs), performance metrics, solution methods (Section 4.1), case studies (Section 4.2), and references.

### 4.1. Review of Methods

In this section, we review methods to solve problems introduced in Section 2.

#### 4.1.1. Dispatching.
Dispatching is the first problem: Given a set of customer requests, the fleet manager needs to match them to AVs. For instance, one can rely on heuristics (38, 102, 103), such as nearest neighbors: From a queue of customer requests, one assigns the nearest AV to each request in a first-come, first-served fashion. Such methods, however, generally lead to suboptimal solutions (81). More structured approaches are based on mathematical programming (see, e.g., 31 and references therein). For instance, in vehicle-centric models, whereby one dispatches $m$ AVs to serve $n$ customer requests, one can introduce binary decision variables $x_{ij} \in \{0, 1\}$ that indicate whether AV $i$ is assigned to customer request $j$. With $r_{ij}$ being the profit of serving customer request $j$ with AV $i$, we get the integer linear program

$$\max_{x_{ij} \in \{0,1\}} \sum_{i=1}^{m} \sum_{j=1}^{n} r_{ij} x_{ij}$$

$$\text{subject to} \sum_{i=1}^{m} x_{ij} \leq 1 \quad \forall j \quad \text{(customers served at most once)}.$$

Despite being an integer linear program, the optimization problem can be solved without integrality constraints, thanks to total unimodularity. Indeed, if an integer linear program is totally unimodular,[2] optimal solutions are integers, even if the integrality constraint is dropped; thus, they

---

[2]A matrix is totally unimodular if the determinant of every square submatrix is 0, −1, or +1. Thus, $\min_x c^\top x$ subject to $Ax = b$ is totally modular if $A$ is totally unimodular and $b$ is an integer.

**Table 1** Methods for analysis and control of autonomous mobility-on-demand systems

| Problems | | | | | | Model[b] | | | | | | | | Solution | | | | | | | |
|---|---|---|---|---|---|---|---|---|---|---|---|---|---|---|---|---|---|---|---|---|---|
| | | | | | | | | T | D | C | | | | Algorithm[e] | | | | | | | |
| Dispatching | Routing | Rebalancing | Ride-sharing | Upkeep[a] | Operational | Network[c] | Invariant | Variant | Stochastic | Deterministic | Exogenous | Endogenous | Constraints[d] | Dispatching | Routing | Rebalancing | Ride-sharing | Upkeep | Performance metrics[f] | Case studies[g] | References |
| ✓ | ✓ | ✓ | ✓ | — | ✓ | V | ✓ | ✓ | ✓ | ✓ | | | 🚗🚗 | ILP | ILP | LP | ILP | — | ⏱, $ | NYC | 38, 132 |
| ✓ | ✓ | ✓ | ✓ | — | ✓ | F | ✓ | ✓ | ✓ | ✓ | | | 🚗🚗 | ← ILP → | | | | — | ⏱, $ | SFO | 73 |
| ✓ | ✓ | ✓ | ✓ | — | ✓ | F | ✓ | ✓ | ✓ | ✓ | | | 🚗🚗 | OA | OA | MILP | OA | — | ⏱ | NYC | 110, 131 |
| ✓ | ✓ | ✓ | ✓ | — | ✓ | F | ✓ | ✓ | ✓ | ✓ | | | 🚗🚗 | ← OA → | | | | — | ⏱, km | AU | 102 |
| ✓ | ✓ | ✓ | ✓ | — | ✓ | V | ✓ | ✓ | ✓ | ✓ | | | 🚗🚗 | ← ILP → | | | | — | ⏱, km, ☆ | MIT | 130 |
| ✓ | ✓ | ✓ | ✓ | 🅿,🔌 | ✓ | F | ✓ | ✓ | ✓ | ✓ | | | 🚗🚗 | ← ILP → | | | | ILP | ⏱ | NYC | 74 |
| ✓ | ✓ | ✓ | ✓ | — | ✓ | C | ✓ | ✓ | ✓ | ✓ | | | 🚗🚗 | ← L → | | | | — | $ | CDU, W | 114 |
| ✓ | ✓ | ✓ | ✓ | — | ✓ | V | ✓ | ✓ | ✓ | ✓ | | | 🚗🚗 | ← OA → | | | | — | ⏱ | SF | 103 |
| ✓ | ✓ | ✓ | ✓ | 🔌 | ✓ | V | ✓ | | ✓ | ✓ | | | 🚗🚗 | ← MILP → | | | | MILP | ⏱, $ | NYC, TO | 123, 124 |
| ✓ | ✓ | ✓ | | — | ✓ | F | ✓ | | ✓ | ✓ | | | 🚗🚗 | ← LP → | | | | — | ⏱, 🚗🚗 | NYC | 55 |
| ✓ | ✓ | ✓ | | — | | F | ✓ | | | ✓ | ✓ | | — | ← QP → | | | | — | SW | NYC, BE | 56 |
| ✓ | ✓ | ✓ | | — | ✓ | F | ✓ | | ✓ | ✓ | | | 🚗🚗 | ← QP → | | | | — | ⏱, $ | SF | 94 |
| ✓ | ✓ | ✓ | | — | ✓ | F | ✓ | | ✓ | ✓ | ✓ | | 🚗🚗 | ← LP → | | | | — | ⏱, $, ☆ | NYC, HZ | 122 |
| ✓ | ✓ | ✓ | | — | ✓ | F | ✓ | | ✓ | ✓ | | | — | ← LP → | | | | — | ⏱, $ | NYC | 72 |
| ✓ | ✓ | ✓ | | 🅿 | ✓ | F | ✓ | | ✓ | ✓ | | | — | ← LP → | | | | LP | ⏱ | AR | 128 |
| ✓ | ✓ | ✓ | | 🔌 | ✓ | F | ✓ | | ✓ | ✓ | | | — | ← LP → | | | | LP | SW, $ | DFW, OC | 53, 54 |
| ✓ | ✓ | ✓ | | — | ✓ | F | ✓ | | ✓ | ✓ | | | 🚗🚗 | ← LP → | | | | — | ⏱, $ | NYC | 101 |
| ✓ | ✓ | ✓ | | — | ✓ | V | ✓ | | ✓ | ✓ | | | 🚗🚗🅿 | ← OA → | | | | — | ⏱, km | CHI | 79 |
| ✓ | ✓ | ✓ | | — | ✓ | F | ✓ | | | ✓ | | ✓ | — | ← MILP → | | | | — | 🚗🚗, $ | HZ | 120 |
| ✓ | ✓ | ✓ | | 🅿 | ✓ | F | ✓ | | ✓ | ✓ | | | 🚗🚗 | ← LP → | | | | LP | ☆ | SF | 69 |
| ✓ | ✓ | ✓ | | — | ✓ | F | ✓ | | ✓ | ✓ | | | 🚗🚗 | ← CO → | | | | — | $ | NYC | 126 |
| ✓ | ✓ | ✓ | | — | ✓ | F | ✓ | | ✓ | ✓ | | | 🚗🚗 | ← LP → | | | | — | $, 🚗🚗 | SG | 129 |
| ✓ | ✓ | ✓ | | — | | F | ✓ | | | ✓ | ✓ | | — | ← LP → | | | | — | SW | NYC | 92 |
| ✓ | ✓ | ✓ | | — | ✓ | F | ✓ | | | ✓ | ✓ | | — | ← CO → | | | | — | ⏱ | NYC | 125 |
| ✓ | ✓ | ✓ | | — | ✓ | F | ✓ | | | ✓ | ✓ | | 🚗🚗 | ← QP → | | | | — | ⏱, $ | NYC | 57 |
| ✓ | ✓ | ✓ | | — | ✓ | F | ✓ | | ✓ | ✓ | | | 🚗🚗 | ← ILP → | | | | — | ⏱, $, km | SG | 95 |
| ✓ | ✓ | ✓ | | 🔌 | ✓ | F | ✓ | | ✓ | ✓ | | | 🚗🚗 | ← MILP → | | | | MILP | ☆ | NB, CA | 76 |
| ✓ | ✓ | ✓ | | — | | Q | ✓ | | ✓ | | ✓ | | 🚗🚗,🔌 | ← LP → | | | | — | ⏱ | NYC | 170 |
| ✓ | ✓ | ✓ | | — | ✓ | — | ✓ | ✓ | | ✓ | | | — | ← L → | | | | — | ⏱, ☆ | NYC | 115 |
| ✓ | ✓ | ✓ | | — | ✓ | V | ✓ | | ✓ | ✓ | | | 🚗🚗 | ← LP → | | | | — | ⏱, $, ☆ | NYC | 121 |
| ✓ | ✓ | | | — | ✓ | C | ✓ | | ✓ | ✓ | | | 🚗🚗 | OA | OA | — | — | — | ⏱ | AR | 84 |
| ✓ | ✓ | | | 🔌 | ✓ | V | ✓ | | ✓ | ✓ | | | 🚗🚗 | QCILP | QCILP | — | — | QCILP | ⏱, km | CO | 75 |
| ✓ | ✓ | | | — | | F | ✓ | | | ✓ | ✓ | | — | L | | L | — | — | ⏱ | LA | 171 |
| ✓ | | ✓ | | — | ✓ | Q | ✓ | | ✓ | ✓ | | | 🚗🚗 | MILP | — | NLP | — | — | ☆ | NY, HZ | 93 |
| ✓ | | ✓ | | — | ✓ | C | ✓ | | ✓ | ✓ | | | 🚗🚗 | OA | — | OA | — | — | ⏱ | SFO | 83 |
| ✓ | | ✓ | | — | ✓ | F | ✓ | | ✓ | ✓ | | | 🚗🚗 | OA | — | OA | — | — | $ | NYC | 71 |
| ✓ | | ✓ | | — | ✓ | F | ✓ | | ✓ | ✓ | | | 🚗🚗 | LP | — | L | — | — | $ | CDU, NYC | 127 |
| ✓ | | ✓ | | — | ✓ | C | ✓ | | ✓ | ✓ | | | 🚗🚗 | OA | — | OA | — | — | ⏱, km | CHI | 172 |
| ✓ | | ✓ | | — | ✓ | V | ✓ | ✓ | | ✓ | | | 🚗🚗 | OA | — | — | OA | — | ⏱ | BE | 173, 174 |
| ✓ | | ✓ | | — | ✓ | C | ✓ | | ✓ | ✓ | | | 🚗🚗 | LP | — | — | MILP | — | ⏱, km, ☆ | NYC | 175 |
| ✓ | | ✓ | | — | ✓ | V | ✓ | | ✓ | ✓ | | | 🚗🚗 | OA | — | — | — | — | $ | AR | 176 |
| ✓ | | | | 🔌 | | Q | ✓ | | ✓ | | ✓ | | 🚗🚗 | NLP | — | — | — | NLP | ⏱ | AR | 96 |
| ✓ | | | | — | ✓ | V | ✓ | | ✓ | ✓ | | | 🚗🚗 | ILP | — | — | — | — | ⏱ | NYC | 77 |
| ✓ | | | | 🔌 | | V | ✓ | | ✓ | ✓ | | | 🚗🚗 | OA | — | — | OA | — | ⏱, $ | NYC | 104 |
| | ✓ | | | — | ✓ | V | ✓ | | ✓ | ✓ | | | 🚗🚗 | — | L | — | L | — | km, ☆ | CO | 105 |
| | ✓ | | | — | ✓ | V | ✓ | | ✓ | | ✓ | | — | — | L | — | — | — | ⏱ | AR | 116 |
| | ✓ | | | — | ✓ | C | ✓ | | ✓ | ✓ | | | — | — | L | — | — | — | ☆ | SG | 117 |
| | | ✓ | | — | ✓ | F | ✓ | | ✓ | ✓ | | | 🚗🚗 | — | — | LP | — | — | ⏱ | AR | 35 |
| | | ✓ | | — | ✓ | Q | ✓ | | ✓ | ✓ | | | — | — | — | LP | — | — | ☆ | NYC | 82 |
| | | ✓ | | — | | F | ✓ | | ✓ | ✓ | | | 🅿 | — | — | MINLP | — | — | $ | SG | 70 |
| | | ✓ | | — | ✓ | C | ✓ | | ✓ | ✓ | | | 🚗🚗 | — | — | OA | — | — | ⏱, ☆ | SFO | 86 |
| | | ✓ | | — | ✓ | F | ✓ | | ✓ | ✓ | | | 🚗🚗 | — | — | LP | — | — | 🚗🚗 | AR | 118 |
| | | ✓ | | — | | Q | ✓ | | ✓ | | ✓ | | 🚗🚗 | — | — | LP | — | — | $ | HZ | 177 |
| | | ✓ | | 🔌 | | F | ✓ | | ✓ | ✓ | | | 🚗🚗 | — | — | NLP | — | NLP | ⏱ | HW, CDU | 68 |
| | | ✓ | | — | ✓ | V | ✓ | | ✓ | ✓ | | | 🚗🚗 | — | — | ILP | — | — | ☆ | NYC | 111 |
| | | ✓ | | — | | F | ✓ | | | ✓ | ✓ | | 🚗🚗 | — | — | MNLP | — | — | ☆ | AR | 178 |
| | | | ✓ | — | | F | ✓ | | | ✓ | ✓ | | 🚗🚗 | — | — | — | LP | — | ⏱, ☆ | SE | 179 |
| | | | ✓ | — | ✓ | Q | ✓ | | | ✓ | ✓ | | 🚗🚗 | — | — | — | LP | — | ☆ | SG | 180 |
| | | | | ✓ | ✓ | F | ✓ | | ✓ | ✓ | | | — | — | — | — | — | OA | $ | AU | 43 |

[a]🔌, solutions for problems related to charging of autonomous vehicles (AVs); 🅿, solutions for scheduling problems related to parking of AVs.
[b]T, time; D, demand; C, congestion.
[c]C, continuum; F, fluidic; Q, queuing-theoretic; V, vehicle-centric.
[d]🚗🚗, fleet size; 🅿, parking; 🔌, limited AV autonomy.
[e]CO, optimization problem with convex cost and constraints; ILP, optimization problem with integer decision variables and linear cost and constraints; L, learning (e.g., reinforcement learning); LP, optimization problem with linear cost and constraints; MILP, optimization problem with integer and real decision variables and linear cost and constraints; MINLP, optimization problem with integer and real decision variables and nonlinear cost and constraints; NLP, optimization problem with nonlinear cost and constraints; OA, own algorithms (e.g., heuristic, provable algorithms); QCILP, optimization problem with integer decision variables, linear cost, and quadratic constraints; QP, optimization problem with quadratic cost and linear constraints.
[f]$, revenue and costs; ⏱, waiting and travel time; km, travel distance; ☆, demand satisfaction; 🚗🚗, fleet size; SW, social welfare.
[g]AR, artificial; AU, Austin; BE, Berlin; CA, California; CDU, Chengdu; CHI, Chicago; CO, Cologne; DFW, Dallas–Fort Worth; HW, hardware; HZ, Hangzhou; LA, Los Angeles; MIT, Massachusetts Institute of Technology campus; NB, Newport Beach; NYC, New York City; OC, Orange County; SE, Seattle; SF, Sioux Falls; SFO, San Francisco; SG, Singapore; TO, Tokyo; W, Wuhan.

can be easily and scalably solved as linear programs. To further improve scalability, a variety of algorithms have been implemented (28–31, 104). Recent works have investigated learning-based approaches (e.g., learning optimal policies via neural networks) to improve real-time capabilities (77, 105).

**4.1.2. Routing.** Routing of AVs is an instance of a dynamic vehicle routing problem, part of the larger family of vehicle routing problems (32). Vehicle routing problems are usually solved (globally or via heuristics) as static routing problems, in which trip origins and destinations are known in advance (for an extensive review, see 106). Crucially, however, in AMoD systems demand is dynamic, leading to dynamic vehicle routing problems. In particular, dynamic vehicle routing problems where people or goods need to be transported between an origin and a destination are dynamic pickup and delivery problems (33), which can be subdivided into three classes: many-to-many problems, in which each request is routed to multiple destinations (107); one-to-one problems, in which each request has a single origin and a single destination (84); and one-to-many problems, in which goods initially located at a depot are delivered to customers, and goods originating from customers are collected and delivered to the depot (108). Clearly, routing AVs is a dynamic one-to-one pickup and delivery problem. However, a full characterization of AMoD systems requires additional properties. First, requests can be immediate and for immediate service, and not meant as prior booking. Second, arrival and travel times are stochastic (16). Finally, apart from ride-sharing cases, AVs are assumed to have single-party occupancy. In AMoD systems, one computes policies (and not routes) to service requests arriving during operations (18, 109). We identify two approaches: custom algorithms (including heuristics and machine learning) (110, 111) and algorithmic queuing theory (18, 112, 113).

Custom algorithms for routing AVs have been studied, for instance, by Jaillet & Wagner (78) and Levin et al. (102). Often, these algorithms are assessed via comparisons with omniscient (noncausal) algorithms or supported by extensive empirical experiments, rarely enjoying theoretic guarantees (102, 114–117). Algorithmic queuing theory usually builds on continuum models of the transportation system, where AVs move at a constant velocity and demand is modeled as a spatiotemporal Poisson process. As optimizing over all policies is intractable, research has focused on designing algorithms, providing rigorous theoretic guarantees on achievable performance, and creating extensions (e.g., to obstacles, time constraints, and priorities) (112, 113). However, most formal performance guarantees for algorithms developed using algorithmic queuing theory hold only in asymptotic regimes, such as a light or heavy load. In medium-load regimes, where systems usually operate, it is difficult to assess the performance of such algorithms. Furthermore, it is hard to incorporate additional operational complexity, such as congestion and charging constraints (81).

**4.1.3. Rebalancing.** AMoD systems tend to become out of balance: When some origins and destinations are more popular than others, AVs build up at some stations and become depleted at others (35). Historically, researchers studied rebalancing for one-way car-sharing schemes (e.g., car2go and Zipcar) and bike-sharing (107, 111). While these systems also suffer from asymmetric demand, they perform infrequent rebalancing (e.g., happening a few times a day). Thus, the continuous operation of AMoD systems calls for new tools to reason about the relocation of customer-free AVs.

In queuing-theoretic settings, rebalancing strategies are computed via linear programs. The resulting real-time implementation is formally an integer linear program, which, however, reduces to a linear program thanks to total unimodularity (93). In network flow models, rebalancing problems are often formulated as (integer) optimization problems (72, 94, 118), possibly considering competitive behaviors (119). For instance, recall the dynamic model in Equation 1. Here, the optimal rebalancing strategy $\alpha_{ij}$ from node $i$ to node $j$ results from the linear program

$$\min_{\alpha_{ij} \geq 0} \sum_{i \in S} \sum_{j \in D} t_{ij} \alpha_{ij}$$

$$\text{subject to} \sum_{j \in D} \alpha_{ij} = -\lambda_i + \sum_{j \neq i} \lambda_j p_{ji} \quad \forall i \in S := \{i \in \{1, \ldots, N\} : \lambda_i < \lambda_1 p_{1i} + \cdots + \lambda_N p_{Ni}\},$$

$$\sum_{i \in S} \alpha_{ij} = \lambda_j - \sum_{i \neq j} \lambda_i p_{ij} \quad \forall j \in D := \{1, \ldots, N\} \setminus S.$$

In most cases, real-time rebalancing has been studied in a model predictive control framework, sometimes combined with stochastic models for future demand, possibly enhanced by machine learning (120, 121). Previous works have also exploited total unimodularity to reduce integer linear programs to linear programs (122).

#### 4.1.4. Dispatching, routing, and rebalancing.
Most works have focused on solving routing, dispatching, and rebalancing problems together, sometimes along with ride-sharing problems, as well as on incorporating specific operational constraints. For instance, in network flow models, we lose the identity of single AVs. Thus, the problems of dispatching, routing, and rebalancing AVs are implicitly solved when computing minimum cost flows, which are usually formulated as linear programs and quadratic programs for noninteger flows (72, 94, 122) and as integer linear programs and mixed-integer linear programs for integer flows (120), possibly considering charging (123, 124) and parking routines (69).

In time-invariant settings, Salazar et al. (56) and Wollenstein-Betech et al. (57) proposed a quadratic program for socially optimal dispatching, routing, and rebalancing, considering interactions with other modes of transportation (e.g., public transport). Among other things, the authors outlined the benefits of cooperatively operating AMoD systems and public transport in New York City and Berlin. Similarly, Rossi et al. (55) and Salazar et al. (92) solved the problems via linear programs, and Solovey et al. (125) did so via convex optimization. Interestingly, the time-invariant setting enables one to account for endogenous congestion effects. With the exception of the paper by Rossi et al. (55), which proposed a real-time implementation based on an integer linear program, these works are meant not to devise operational policies but to enable planning studies.

In time-varying settings, the problems are often cast within a model predictive control framework: Carron et al. (94) formulated a quadratic program, and Kang & Levin (69), Zgraggen et al. (72), and Tsao et al. (122) formulated a linear program, sometimes accounting for parking. Similarly, Boewing et al. (76), Zhang et al. (123), and Iacobucci et al. (124) framed the problem as a mixed-integer linear program, considering the charging of electric AVs. Ma et al. (101) proposed a linear program, and Yang et al. (95) proposed a model predictive control framework based on Stackelberg games, solved via a linear program. Miao et al. (126) adopted a distributionally robust approach with respect to uncertainty in future demand, solving the problem via convex optimization, and Guériau et al. (115) offered a learning-based solution that uses neural networks, together with microscopic simulations of the transportation system, to learn optimal policies. All of these approaches are meant for real-time implementation, possibly up to relaxations, heuristics, or suboptimality of the underlying optimization problems. Heuristic algorithms were proposed by Hyland & Mahmassani (79). Finally, Gammelli et al. (127) decoupled dispatching and rebalancing (neglecting routing) in two linear programs and linked them through reinforcement learning: Once AVs are dispatched by solving the first linear program, the optimal distribution of customer-free AVs is computed via reinforcement learning (based on graph neural networks) and realized by solving the second linear program.

Other optimization-based approaches have been proposed for planning studies (128, 129), possibly in combination with the power grid (53, 54). In these studies, for instance, the authors showed that for a case study in Orange County, California, coordination between AVs and the power grid eliminated 99% of overloads and 50% of voltage drops.

### 4.1.5. Dispatching, routing, and rebalancing with ride-sharing.

Tsao et al. (73) framed dispatching, routing, rebalancing, and ride-sharing as network flow problems and formulated a mixed-integer linear program to optimize the costs and times of operations, implemented in real time in a receding-horizon fashion that relies on forecasting for the demand. By contrast, Miller & How (130) relied on integer programming and, to overcome the computational burden, heuristics. Alonso-Mora et al. (38), Sayarshad & Chow (110), and Ma et al. (131) broke the problem down into two instances. Alonso-Mora et al. (38) solved dispatching, routing, and ride-sharing via an integer linear program and solved rebalancing through a linear program. Among other things, the authors showed that, without ride-sharing, 98% of the taxi rides in New York City, currently served by more than 13,000 taxis, could be served with just 3,000 AVs, and that ride-sharing leads to substantial additional benefits (e.g., less travel distance). Fielbaum et al. (132) extended the work of Alonso-Mora et al. (38) to assess the benefits of including walking sections to maximize system performance. Sayarshad & Chow (110) and Ma et al. (131) solved dispatching, routing, and ride-sharing via an ad hoc heuristic algorithm and solved rebalancing via a mixed-integer linear program, whose solutions are computed approximately via Lagrange decomposition. Levin et al. (102) and Liu et al. (103) relied entirely on ad hoc algorithms, and Lin et al. (114) relied on reinforcement learning. Finally, Chu et al. (74) combined graph-theoretic and vehicle-centric models to formulate the problem as an integer linear program, including the dynamic state of charge of the electric AVs and parking.

### 4.2. Existing Tools for Simulation

There are two types of simulation tools for AMoD systems: custom-made and general transportation simulators. While custom-made simulators include ad hoc software implementations to numerically analyze and control AMoD systems, they are typically not adaptable to novel scenarios, they do not generalize, and, in some cases, they are simplistic. General transportation simulators, by contrast, entail greater software complexity and result from years of development and validation (133–135). Simulators can be further divided into three classes: microscopic, macroscopic, and mesoscopic. Microscopic simulators model both AVs and traffic dynamics (136) and are commonly used when modeling microscopic effects such as navigation of intersections or merging on freeways (135, 137–139). Macroscopic simulators are designed to analyze effects at a higher abstraction level (140); transportation systems are generally abstracted as coarse-grained graphs, and agents are represented as flows. Mesoscopic simulators operate at an intermediate abstraction level, omitting the dynamics of individual AVs but maintaining a faithful representation of the network, demand, and fleet size (45, 56, 141, 142).

Researchers have identified three ways to validate algorithms: synthetic, hardware, and real-world case studies. Synthetic case studies rely on the generation of artificial networks and demand and allow for quick prototyping and proofs of concept (e.g., scalability). Hardware tests provide more practical observations but are feasible only on small scales (82, 143, 144). For these reasons, most works have presented real-world case studies, usually constructed as follows. Road networks can be downloaded from OpenStreetMap (145), and schedules of public transportation systems are often publicly available as General Transit Feed Specification (GTFS) data (146). Hence, the availability of demand data drives the design of the case study, and, in particular, the choice of city.³ New York City has been a pioneer, releasing taxi demand data via the New York City Taxi and Limousine Commission every year since 2009. The data sets include 697,622,444 trips (147). Thus, the vast majority of the studies in **Table 1** are based in New York City, but more recently, case studies have also been performed for numerous other cities.

## 5. FUTURE RESEARCH AVENUES

We identify four main avenues for future research: modeling, co-design of AV-enabled transportation systems, interactions among stakeholders in the mobility ecosystem, and fairness, privacy, and trust.

### 5.1. Modeling

The need for more efficient analysis and control of AMoD systems calls for richer models in terms of demand, congestion, operational constraints, and ride-sharing.

**5.1.1. Elastic demand.** So far, researchers have considered mainly inelastic demand. The study of elastic demand for AMoD systems is therefore a natural direction for future research. This is an interdisciplinary research area, touching upon topics such as incentives, customer preferences, and mode choice (51, 63, 99, 148, 149). Questions of interest include how AMoD systems can attract new customers, what effective tools a mobility provider can use to steer mode choice, and whether AMoD systems are scalable to accommodate induced demand.

**5.1.2. Congestion.** As suggested by **Table 1**, most existing operational policies ignore endogenous congestion. While in the last few years more research in this direction has been pursued [e.g., in the seminal work by Rossi (150)], more questions need to be answered, mainly along two directions: (*a*) approaches to reduce congestion, including ride-sharing, demand staggering, and integration with other mobility systems (e.g., public transport and micromobility), and (*b*) computationally efficient methods for more accurate congestion models (125).

**5.1.3. Operational constraints.** While considerable attention has been dedicated to fleet size and charging constraints, other important aspects have been ignored. We identify two main research gaps.

*5.1.3.1. Autonomy-related limits.* First, high usage of AVs requires not only scheduling recharging or refueling sessions but also considering autonomy-related operational limits, such as energy consumption resulting from intense sensing and computing [autonomy is estimated to increase a vehicle's energy consumption by 20% (151)], and (autonomous) calibration (58, 152). Second, one needs to strategically locate (autonomous) calibration facilities and efficiently schedule calibration sessions. Facilities could also be used for general AV maintenance and cleaning. Indeed, in an AV, no driver can ensure that customers leave the vehicle clean and ready for the next customers.

---

³We have created a list of existing publicly available demand data sets for the simulation of AMoD systems; see **http://gioele.science/amod-demand-data**.

***5.1.3.2. Autonomy-enabling infrastructure.*** In this article, we have tacitly assumed that automating vehicles is the only feasible way to achieve AMoD. Yet this is just part of the story. One interesting alternative is autonomy-enabling infrastructure, in which part of the autonomy pipeline of AVs is outsourced to the public infrastructure [e.g., online, cloud-based perception and mapping (59, 153)]. Open questions include whether this approach is feasible, how much it would cost, and whether it would scale effectively.

**5.1.4. Ride-sharing.** Various studies have outlined potential benefits and challenges of ride-sharing (36–38, 154). While recent studies are a first step in the systematic assessment of ride-sharing (36), further work needs to be done to determine to what extent ride-sharing is beneficial and whether it is reasonable only for certain urban environments (e.g., only for specific network topologies).

### 5.2. Co-Design of Autonomous Vehicle–Enabled Transportation Systems

The current design process for future mobility solutions lacks clear, specific requirements for the service they would be providing (14). Indeed, knowledge about the service could drastically impact their design. For instance, knowing that an AV does not need to reach speeds higher than 30 mph (e.g., due to slow average speeds in cities) might significantly simplify its design and accelerate its deployment (39, 45). Furthermore, from the system-level perspective of transportation management, knowledge about the development of new mobility approaches has an impact on future infrastructure investments and provisions of service. Future mobility solutions need to be co-designed with the transportation system they enable. For the particular case of AMoD systems, this requires tools to study their design all the way from the autonomy of AVs to their effects on fleet control and on the transportation system at large (46, 155). Existing approaches for such coupled problems typically have a fixed, problem-specific structure and therefore do not allow one to rigorously design the transportation infrastructure in a modular and compositional fashion (45, 156, 157).

### 5.3. Interactions Among Stakeholders in the Mobility Ecosystem

The rise of private, profit-maximizing mobility providers leveraging public resources, such as ride-hailing companies and, in the near future, AMoD systems, raises questions for current regulation schemes, such as whether mobility providers should be taxed for usage of public resources and how many AVs a city should allow. These questions call for tools to reason about interactions among stakeholders in the mobility ecosystem, including municipalities, mobility providers, and customers. First, municipalities need to design regulatory policies tailored to new mobility options, such as limitations on fleet sizes, public transport prices, and taxes (51, 63–65, 158, 159). Second, different mobility providers will operate in the same environment, opening the field for competitive analyses (64, 65, 160, 161). Third, new mobility approaches will influence travel needs. Finally, coordination schemes for AMoD systems could relate to transportation of goods. Examples include delivery of packages (47) and shipping (162).

### 5.4. Fairness, Privacy, and Trust

The penetration of AMoD systems calls for considerations of fairness, privacy, and trust. First, fairness is an important aspect for centrally controlled systems; for example, the discrepancy among users' travel times can reach 100% in real-world transportation networks (163).

Importantly, stakeholders need to find the right trade-off between fairness and efficiency in the transportation system and study incentive mechanisms to steer mobility providers accordingly (164, 165). Furthermore, while customers' data can be used to improve the service quality of transportation systems, privacy and trust concerns might arise (166–168). Privacy includes handling methodologies for data containing sensitive information, and trust concerns appear as soon as mobility providers and regulation authorities have misaligned interests (167, 169).

## 6. CONCLUSION

In this review, we have provided a full picture of the analysis and control of AMoD systems by identifying relevant problem settings, giving an overview of mathematical models, and analyzing existing solution approaches, highlighting results, trends, and limitations. Finally, we have armed the reader with insights on future research.

### SUMMARY POINTS

- Problems: The literature on the analysis and control of autonomous mobility-on-demand (AMoD) systems focuses on two classes of problems: operational policies [e.g., how to route autonomous vehicles (AVs)] and strategic planning (e.g., where to place charging stations).
- Mathematical models: Modeling AMoD systems requires models of the transportation network, AVs, demand, congestion, operational constraints, and interactions with existing infrastructure.
- Solution approaches: Researchers have developed a wide variety of methods to solve problems related to AMoD systems, relying on mathematical optimization, custom algorithms, and machine learning. Performance metrics include costs, waiting and travel time, travel distance, demand satisfaction, fleet size, and social welfare.

### FUTURE ISSUES

- Models: Effects such as elastic demand, endogenous congestion, operational constraints, and ride-sharing require further investigation.
- Co-design: Mobility solutions need to be co-designed with the system they enable.
- Interactions: We need tools to rigorously reason about interactions among stakeholders in the mobility ecosystem.
- Fairness, privacy, and trust: Deployment of AMoD raises concerns regarding fairness among customers as well as privacy and trust when handling private data.

## DISCLOSURE STATEMENT

AS05CH18_Frazzoli    ARjats.cls    November 13, 2021    16:48## ACKNOWLEDGMENTS

The authors' research is supported by the Swiss National Science Foundation under National Centre of Competence in Research (NCCR) Automation (grant agreement 51NF40_180545) and the National Science Foundation [Faculty Early Career Development Program (CAREER) award CMMI1454737]. The authors thank Sonia Monti for graphics support and Devansh Jalota for useful discussions.
## LITERATURE CITED

1. UN Dep. Econ. Soc. Aff. 2021. 68% of the world population projected to live in urban areas by 2050, says UN. *United Nations Department of Economic and Social Affairs*, May 16. https://www.un.org/development/desa/en/news/population/2018-revision-of-world-urbanization-prospects.html
2. UN. 2021. Cities and pollution. *United Nations*. https://www.un.org/en/climatechange/climate-solutions/cities-pollution
3. Schrank D, Eisele B, Lomax T. 2019. *Urban mobility report 2019*. Rep., Tex. A&M Transp. Inst., College Station
4. Mitchell WJ, Borroni-Bird CE, Burns LD. 2010. *Reinventing The Automobile: Personal Urban Mobility for the 21st Century*. Cambridge, MA: MIT Press
5. Pavone M. 2015. Autonomous mobility-on-demand systems for future urban mobility. In *Autonomes Fahren: technische, rechtliche und gesellschaftliche Aspekte*, ed. M Maurer, J Gerdes, B Lenz, H Winner, pp. 399–416. Berlin: Springer
6. City of New York. 2021. *OneNYC 2050: building a strong and fair city*. Rep., City of New York. http://onenyc.cityofnewyork.us/reports-resources/
7. Berger T, Chen C, Frey CB. 2018. Drivers of disruption? Estimating the Uber effect. *Eur. Econ. Rev.* 110:197–210
8. Rogers B. 2015. The social costs of Uber. *Univ. Chicago Law Rev. Online* 82:85–102
9. AAA Found. Traffic Saf. 2020. *New American driving survey: updated methodology and results from July 2019 to June 2020*. Tech. Rep., AAA Found. Traffic Saf., Washington, DC. https://aaafoundation.org/new-american-driving-survey-updated-methodology-and-results-from-july-2019-to-june-2020
10. Hancock PA, Nourbakhsh I, Stewart J. 2019. On the future of transportation in an era of automated and autonomous vehicles. *PNAS* 116:7684–91
11. Becker H, Becker F, Abe R, Bekhor S, Belgiawan PF, et al. 2020. Impact of vehicle automation and electric propulsion on production costs for mobility services worldwide. *Transp. Res. A* 138:105–26
12. Iagnemma K. 2020. 100,000 self-driving rides strong. *Aptiv*, Feb. 11. https://www.aptiv.com/en/insights/article/100000-Self-Driving-Rides-Strong
13. Schwall M, Daniel T, Victor T, Favaro F, Hohnhold H. 2020. Waymo public road safety performance data. arXiv:2011.00038 [cs.RO]
14. Yigitcanlar T, Wilson M, Kamruzzaman M. 2019. Disruptive impacts of automated driving systems on the built environment and land use: an urban planner's perspective. *J. Open Innov. Technol. Mark. Complex.* 5:24
15. Kim G, Ong YS, Heng CK, Tan PS, Zhang NA. 2015. City vehicle routing problem (city VRP): a review. *IEEE Trans. Intell. Transp. Syst.* 16:1654–66
16. Psaraftis HN, Wen M, Kontovas CA. 2016. Dynamic vehicle routing problems: three decades and counting. *Networks* 67:3–31
17. Eksioglu B, Vural AV, Reisman A. 2009. The vehicle routing problem: a taxonomic review. *Comput. Ind. Eng.* 57:1472–83
18. Bullo F, Frazzoli E, Pavone M, Savla K, Smith SL. 2011. Dynamic vehicle routing for robotic systems. *Proc. IEEE* 99:1482–504
19. Gendreau M, Potvin JY. 1998. Dynamic vehicle routing and dispatching. In *Fleet Management and Logistics*, ed. TG Crainic, G Laporte, pp. 115–26. Berlin: Springer
*www.annualreviews.org* • *Analysis and Control of AMoD Systems*    18.19